# A New Layered Kagome Strip Structure Na$_2$Co$_3$(AsO$_4$)$_2$(OH)$_2$: Static and Dynamic Magnetic Properties

Duminda S. Liurukara[a], Emily D. Williams[b], Tianran Chen[c], Stuart Calder[a], V. Ovidiu Garlea[a], C. Charlotte Buchanan[d], Dustin A. Gilbert[c,d], Joseph W. Kolis[b], D. A. Tennant[d]

[a]Neutron Scattering Division, Oak Ridge National Laboratory, Oak Ridge, TN 37831, USA
[b]Department of Chemistry and Center for Optical Materials Science and Engineering Technologies (COMSET), Clemson University, Clemson, SC 29634-0973, USA
[c]Department of Physics and Astronomy, University of Tennessee, Knoxville, TN 37996, USA
[d]Department of Materials Science and Engineering, University of Tennessee, Knoxville, TN 37996, USA

**Abstract**

One-dimensional kagome strip chains share much of the same frustrated structural motif as two-dimensional kagome antiferromagnets, making them valuable for deepening our understanding of kagome lattice magnetism. In this paper, we report the hydrothermal synthesis and detailed structural and property characterization of Na$_2$Co$_3$(AsO$_4$)$_2$(OH)$_2$, a striped kagome system. The crystal structure was characterized using single crystal X-ray diffraction which reveals that Na$_2$Co$_3$(AsO$_4$)$_2$(OH)$_2$ crystallizes in monoclinic crystal system *C*2/*m*. The structure features a one-dimensional kagome strip lattice built from Co$^{2+}$ ions and undergoes an antiferromagnetic transition at $T_N$ = 14 K. The magnetic ground state at zero field was characterized using neutron powder diffraction. Below the magnetic transition, Na$_2$Co$_3$(AsO$_4$)$_2$(OH)$_2$ orders into an antiferromagnetic structure with a ***k***-vector (0.5,0.5,0.5). In the proposed model, the Co1 moment is predominantly confined to the *ac*-plane while the Co2 moment is primarily aligned along the *b*-axis. Two flat bands were observed in the inelastic neutron spectra below the magnetic transition, 5 and 10 meV. Inelastic neutron spectra were modeled with a Heisenberg Hamiltonian including three nearest-neighbor exchange interactions ($J_1$, $J_2$, $J_3$) and strong single-ion anisotropy to stabilize the observed magnetic structure. Our study highlights the complexity of $Co^{2+}$-based kagome strip magnetic lattice compound $Na_2Co_3(AsO_4)_2(OH)_2$ which provides an excellent platform to broaden our understanding of the frustrated kagome magnetic lattice space.

**Introduction**

One recent area of research for solid-state chemists has been the study of frustrated magnetic materials, as they can host exotic magnetic ground states that are key to future potential quantum applications such as quantum computing and information processing.[1-4] For example materials chemists can design geometrically frustrated solids with trigonal symmetry that can lead to exotic magnetic ground states such as quantum spin liquids (QSL). In QSL materials, no long-range magnetic ordering appears at low temperatures due to the strong quantum fluctuations of magnetic moments caused by highly degenerate ground states. This can lead to broad continuous magnetic excitations which can be observed in inelastic neutron experiments. This behavior is believed to be associated with fractionalized quasiparticles known as spinons.[5-7] It is very difficult to unambiguously observe QLSs but several two-dimensional (2D) and three-dimensional (3D) magnetic materials have been proposed as candidates.[8-11] The 2D materials containing trigonal symmetry, such as kagome, triangular and honeycomb magnetic materials, have been investigated extensively as candidates for QSLs because they can display extensive magnetic frustration,[12-15] since nearest-neighbor antiferromagnetic interactions cannot be simultaneously satisfied.[14-15] On the other hand, honeycomb magnetic materials, though not geometrically frustrated, can also potentially exhibit a QSL state due to bond-dependent interactions as described in the Kitaev model.[16-17]

In general, a small spin value ($S = 1/2$) is essential to facilitate the quantum fluctuations in these materials.[5-6,8] Examples are Cu-based $S = 1/2$ triangular materials such as CsCu$X_4$ ($X$ = Cl and Br)[18-19] and κ-(BEDT-TTF)$_2$Cu$_2$(CN)$_3$, as well as Yb$^{3+}$ (Kramers doublets with an effective spin $S_{eff} = 1/2$) based triangular magnetic materials such as YbMgGaO$_4$ and $A$Yb$Q_2$ ($A$ = Na, K, Cs; $Q$ = O, S, Se).[20-21] A confluence of features has driven a significant research effort into the study of spin-1/2 kagome lattice antiferromagnetic materials such as Herbertsmithite ZnCu$_3$(OH)$_6$Cl$_2$,[22] α-vesignieite BaCu$_3$V$_2$O$_8$(OH)$_2$[23] and [NH$_4$]$_2$[C$_7$H$_{14}$N][V$_7$O$_6$F$_{18}$]$_5$.[23] The corner-sharing triangles in kagome materials create strong magnetic frustration, making the spin-1/2 kagome system one of the most promising candidates for realizing a QSL state. Intriguingly, these exotic magnetic phases have also been proposed in another class of kagome-type magnetic materials, namely, kagome strips.[27-30] The kagome strip magnetic lattice is analogous to the kagome lattice where both magnetic lattices consist of 1D kagome strip chains as building blocks. Moreover, these materials can often exhibit unusual behavior in the presence of external magnetic fields. One characteristic behavior for example, is for an $S = 1/2$ 2D kagome lattice to exhibit multiple magnetization plateaus at fractional values of the saturated magnetic moment when plotted against external field.[25-27] Theory predicts that these magnetic plateaus may be classical or quantum in nature.[27-30]

Despite their promising behavior, however, only a handful of kagome strip compounds have been synthesized and studied in detail and the bulk of these contain only $S = 1/2$ Cu$^{2+}$ ions. Recently, Tang et al. synthesized two $S = 1/2$ kagome strip structures, $A_2$Cu$_5$(TeO$_3$)(SO$_4$)$_3$(OH)$_4$ ($A$ = Na, K) using a low-temperature hydrothermal method.[31] These two compounds possess a highly distorted kagome strip magnetic lattice with three different Cu$^{2+}$ crystallographic sites.

Both of these compounds exhibit antiferromagnetic behavior with Weiss temperatures of $\theta$ = -6.1 K and -13.9 K for Na and K, respectively. Following this first report, theoretical investigations of $S = 1/2$ kagome strip lattices have generated significant interest in modeling the magnetic phase diagram of these systems.[31] A recent paper from Morita et al. presents a detailed density-matrix renormalization-group (DMRG) investigation of the magnetic phase diagram of $A_2Cu_5(TeO_3)(SO_4)_3(OH)_4$ in the presence of an applied magnetic field.[28] Their discovery confirms the presence of wide variety of magnetizations plateaus. Despite these studies, experimental verification of anisotropic magnetic properties has not yet been possible due to the inaccessibility of suitably large single crystals. Another interesting $S = 1/2$ kagome strip material was reported by Kakarla et al., α-$Cu_5O_2(SeO_3)_2Cl_2$ made using chemical vapor transport method.[32] This material contains a highly distorted kagome strip lattice that orders antiferromagnetically at $T_N$ = 35 K and exhibits a broad maximum in magnetic susceptibility, indicating short-range ordering due to the low dimensionality. The magnetic structure of α-$Cu_5O_2(SeO_3)_2Cl_2$ was characterized using neutron powder diffraction, but no further investigation was performed using inelastic neutron scattering to investigate the short-range ordering observed in the magnetic susceptibility.[32] Another interesting kagome strip material is $Ca_2Mn_3O_8$.[33] This material was only synthesized in powder form and exhibited an antiferromagnetic transition observed at $T_N$ = 58 K. Neutron powder diffraction was employed to characterize the magnetic structure of $Ca_2Mn_3O_8$ which is notably different from α-$Cu_5O_2(SeO_3)_2Cl_2$. The neutron diffraction results indicate that the Mn moments in $Ca_2Mn_3O_8$ arrange perpendicular to the kagome strip plane as two-up-two-down, i.e. ↑↑↓↓ spin arrangement. These intriguing magnetic properties have prompted us to explore novel 2D kagome strip magnetic lattices and investigate the underlying physics proposed by recent theoretical studies.

Another interesting but comparatively less explored class of kagome strip structures belongs to compounds with the general formula $A_2M_3(EO_4)_2(X)_2$ ($A$ = alkali metal; $M$ = $Mn^{2+}$, $Co^{2+}$, $Ni^{2+}$; $E$ = V, P, As; $X$ = OH, F). Over the past several years, three isostructural members of this family have been reported by us and others, namely $Na_2Ni_3(PO_4)_2(OH)_2$, $K_2Mn_3(VO_4)_2(OH)_2$, and $Na_2Co_3(VO_4)_2(OH)_2$.[34-37] All these compounds crystallize in the monoclinic crystal system with space group $C_2/m$ and host a two-dimensional kagome strip magnetic lattice. $Na_2Ni_3(PO_4)_2(OH)_2$ exhibits an antiferromagnetic transition at $T_N$ = 38.4 K, accompanied by a characteristic low-dimensional hump in the magnetic susceptibility. Notably, both $K_2Mn_3(VO_4)_2(OH)_2$ and $Na_2Ni_3(PO_4)_2(OH)_2$[34] were initially synthesized as small crystals using mild hydrothermal conditions (≈200 °C). To date, however, detailed magnetic and neutron-scattering studies of this family remain lacking, providing ample opportunities for materials chemists to uncover novel and potentially exotic magnetic states associated with kagome strip structures.

Nevertheless, the synthesis of novel magnetic materials with well-controlled chemical composition and minimal structural disorder remains a major challenge, particularly for pyrochlore and kagome-based systems. This limitation has hindered the unambiguous identification of QSL behavior in pyrochlore and kagome materials, which are highly sensitive to subtle chemical disorder and lattice imperfections. Common sources of perturbation include metal-site mixing, such as A/B-site disorder in pyrochlores or Zn/Cu site disorder in herbertsmithite, as well as deviations from ideal stoichiometry arising from intrinsic lattice defects in oxides.[38-40] Because many of these materials are synthesized as polycrystalline powders via high-temperature solid-state or melt-based routes, achieving precise control over site occupancy and defect concentrations is often difficult using conventional preparative methods. To overcome these challenges, we developed a high-temperature hydrothermal synthesis strategy that enables improved control over cation ordering and defect formation. In addition, this approach frequently yields large, high-quality single crystals, providing materials that are well suited for detailed structure–property investigations, including single-crystal magnetic measurements and neutron scattering.[41-42] Thus, we applied our high-temperature hydrothermal synthesis approach to the targeted growth of large single of novel kagome-strip frameworks. Using this strategy, we have recently expanded the $A_2M_3(EO_4)_2(X)_2$ materials family through the synthesis of several new compositions. The chemical flexibility of this framework provides a versatile platform for systematically exploring the effects of metal-ion substitution, spin state, and bridging-group chemistry, thereby enabling fine control of magnetic exchange interactions and structure–property relationships.

Here we present the synthesis and detailed structural characterization of single crystals of $Na_2Co_3(AsO_4)_2(OH)_2$, as well as a comprehensive physical property characterization using both bulk probes and neutron scattering techniques. The crystal structure investigation displays a 2D $Co^{2+}$ kagome strip magnetic lattice with two crystallographically unique $Co^{2+}$ ions. Magnetic properties were determined using ground single crystals and display an antiferromagnetic transition at $T_N$ = 14 K. We used neutron powder diffraction to characterize the ordered zero-field magnetic structure and demonstrate that the ground state is an antiferromagnetic structure with a propagation vector **k** = (0.5,0.5,0.5) which doubles the magnetic unit cell in all three crystallographic directions. Inelastic neutron scattering shows that the spin wave spectrum possesses two magnetic excitations, which can be explained by using three nearest neighbor interactions and single ion anisotropy. Our findings suggest that $Na_2Co_3(AsO_4)_2(OH)_2$ is an attractive candidate to study the low dimensional magnetism associated with the kagome strip magnetic lattices.

**Experimental Section**

**Single crystal growth of $Na_2Co_3(AsO_4)_2(OH)_2$**

$Na_2Co_3(AsO_4)_2(OH)_2$ was synthesized via a high-temperature/high-pressure hydrothermal technique using Tuttle-seal autoclaves. In a typical reaction, a total mixture of 0.2 g of CoO and $As_2O_5$ were mixed using a 2:1 molar ratio with 0.4 mL of 5 M NaOH mineralizer. The reactants and the mineralizer were loaded to silver ampoules (2.5 inch long and ¼ inch outer diameter). The ampoules were welded shut and loaded into the autoclave with water to provide counter pressure. Reactions were run for 7 days at 500 °C and then cooled to room temperature. Thin column shaped (~0.5-1 mm in length) single crystals were recovered by washing with DI water and acetone. The chemicals used in this study: CoO (Thermo Scientific, 95%), $As_2O_5$ (Thermo Scientific, 99.9%), and NaOH (Acros Organics, 98%).

**Characterization**

The crystal structure of $Na_2Co_3(AsO_4)_2(OH)_2$ was determined by using single crystal X-ray diffraction on a Bruker D8 Venture diffractometer. Data were collected at room temperature utilizing Mo Kα radiation, λ = 0.71073 Å, integrated and corrected for absorption SAINT, SADABS available in the APEX 3 software.[44] Structure refinements were performed using the SHELXTL software.[45] All non-hydrogen atoms were refined anisotropically while hydrogen atoms were found from the difference electron density map. The phase purity of the synthesized samples was checked using powder diffraction. Elemental analysis of single crystals was performed by energy dispersive X-ray analysis (EDX) on a Hitachi SU6600 VP FE-SEM with a 20 kV accelerating voltage. The unit cell parameters and fractional atomic coordinates with equivalent isotropic displacement parameters are shown in Table 1 and 2, respectively. The selected bond distances and angles of $Na_2Co_3(AsO_4)_2(OH)_2$ are displayed in Table 3.

Magnetic properties were determined using ground single crystals with a mass of 20 mg. Magnetometry measurements were performed using a vibrating sample magnetometer attached to the Quantum Design Dynacool PPMS. Temperature dependent magnetic measurements were collected from 350 to 2 K with the applied magnetic field ranging from 100 Oe to 90 kOe. Isothermal magnetization data were collected up to 90 kOe from 2 to 100 K.

Neutron powder diffraction (NPD) measurements were performed using the HB2A Powder Diffractometer at High Flux Isotope Reactor (HFIR) at Oak Ridge National Laboratory.[46] Diffraction patterns were collected using constant wavelengths of 2.41 Å from the Ge(113) monochromator reflection and 1.54 Å from the Ge(115) reflection. The neutron diffraction data were analyzed by using the FullProf Suite Package.[47] Symmetry-allowed magnetic structures were considered using both representational analysis with *SARAh* and magnetic space groups with the Bilbao Crystallographic Server.[48-50]

Inelastic-neutron-scattering (INS) measurements were performed using the HYSPEC spectrometer at Spallation Neutron Source in Oak Ridge National laboratory. The data were collected with the incident energies $E_i$ = 50 and 20 meV and the Fermi chopper frequency of 360 Hz. These measurements were performed using the same ground single crystal sample that was used for the diffraction experiment. The INS data reduction and visualization were done with the MANTID software package and spin-wave calculations were performed using **Sunny.jl**.[51]

**Table 1** Crystallographic data of $Na_2Co_3(AsO_4)_2(OH)_2$ determined by single crystal X-ray diffraction.

| Empirical Formula | $Na_2Co_3As_2O_{10}H_2$ |
|---|---|
| Formula Weight (g/mol) | 534.63 |
| Crystal System | monoclinic |
| Crystal Dimensions (mm) | 0.08 x 0.19 x 0.21 |
| Space Group, $Z$ | $C2/m$ (No. 12), 2 |
| Temperature (K) | 298 |
| $a$ (Å) | 14.5885(9) |
| $b$ (Å) | 5.9376(3) |
| $c$ (Å) | 5.0992(3) |
| $\beta$ (°) | 103.63(2) |
| $V$ (Å$^3$) | 429.26(4) |
| $d_{calc}$ (g/cm$^3$) | 4.136 |
| $\lambda$ (Å) | 0.71073 |
| $\mu$ (mm$^{-1}$) | 13.525 |
| $T_{min}$ | 0.1620 |
| $T_{max}$ | 0.4150 |
| $\theta$ range (°) | 4.43-26.49 |
| No. of Reflns. | 2113 |
| Unique Reflns. | 428 |
| Observed Reflns. | 1347 |
| No. of Parameters | 53 |
| No. of Restraints | 0 |
| final R [$I > 2\sigma(I)$] R1, wR2 | 0.0125, 0.0319 |
| final R (all data) R1, wR2 | 0.0125, 0.0319 |
| GOF | 1.225 |
| Largest diff. peak/hole, e/Å$^3$ | 0.317/-0.324 |

**Table 2** Fractional atomic coordinates and isotropic displacement parameters (Å$^2$) of Na$_2$Co$_3$(AsO$_4$)$_2$(OH)$_2$ obtained from single crystal X-ray diffraction

| Atom | Wyckoff | x | y | z | $U_{eq}$ |
|---|---|---|---|---|---|
| Na | 4i | 0.2715(1) | 0 | -0.2958(3) | 0.0280(4) |
| Co1 | 2b | 0.50000 | 1.00000 | 1.0000 | 0.0081(2) |
| Co2 | 4h | 0.5000 | 0.7356(7) | 0.5000 | 0.0080(2) |
| As | 4i | 0.3741(2) | 0.5000 | -0.0902(6) | 0.0064(2) |
| O1 | 4i | 0.5697(1) | 1.0000 | 0.7131(4) | 0.0089(4) |
| O2 | 4i | 0.4077(1) | 0.5000 | 0.2539(4) | 0.0104(4) |
| O3 | 8j | 0.4120(1) | 0.2645(2) | -0.2213(3) | 0.0108(3 |
| O4 | 4i | 0.2567(1) | 0.5000 | -0.1881(5) | 0.0208(6) |
| H1 | 4i | 0.6280(4) | 1.0000 | 0.7450(9) | 0.0290(12 |

**Table 3** Selected bond distances (Å) and angles (°) of Na$_2$Co$_3$(AsO$_4$)$_2$(OH)$_2$.

| Co1O$_6$ | | Co2O$_6$ | |
|---|---|---|---|
| Co1–O1 × 2 | 1.9702(1) | Co2–O1 × 2 | 2.0407(1) |
| Co1–O3 × 4 | 2.1668(1) | Co2–O3 × 2 | 2.1264(1) |
| | | Co2–O2 × 2 | 2.1310(1) |
| As1O$_4$ | | | |
| As1–O4 | 1.6633(2) | Co1–O1–Co2 | 96.63(3) |
| As1–O3 × 2 | 1.7001(1) | Co1–O3–Co2 | 88.34(3) |
| As1–O2 | 1.7079(1) | Co2–O1–Co2 | 100.65(3) |
| | | Co2–O2–Co2 | 82.01(2) |
| Co1–Co2 | 2.994(2) | | |
| Co2–Co2 | 2.798(3) | | |
| Co2–Co2 | 3.140(2) | | |

**Results and Discussion**

**Crystal Structure of Na$_2$Co$_3$(AsO$_4$)$_2$(OH)$_2$**

Hydrothermally grown Na$_2$Co$_3$(AsO$_4$)$_2$(OH)$_2$ single crystals were recovered as pink thin columns with an average size of 0.5-1 mm in length, Figure 1a. The compound crystallizes in the monoclinic crystal system with a space group of $C2/m$ (No.12). The unit cell parameters are $a$ = 14.5885(9) Å, $b$ = 5.9376(3) Å, $c$ = 5.0992(3) Å, $β$ = 103.63(2)° and $V$ = 429.26(4) Å$^3$. There are two crystallographic Co sites, one sodium (Na) site, one arsenic (As) site, and four oxygen (O) sites. We did not observe any detectable disorder in the structure by X-ray diffraction. All crystallographic parameters are given in Table 1 and Table 2. The two Co sites are 2*b* (0.5, 1.0, 1.0) and 4*h* (0.5, 0.7356, 0.5) Wyckoff sites, respectively and both form highly distorted $CoO_6$-pseudo-octahedra (*oct*). As shown in polyhedral view in Figure 1b, $CoO_6$-*oct* form a Co–O–Co 2D layers along the *bc*-plane while the $AsO_4$ tetrahedra connect with the Co–O–Co 2D layers by isolating them along the *a*-axis. Figure 1c displays the similar connectivity along the *c*-axis. Additionally, $Na^+$ ions occupy the gaps between the Co–O–As layers (Figure 1b, c) and this provides complete isolation for the 2D Co–O–Co magnetic layers which enhances the low-dimensionality of $Na_2Co_3(AsO_4)_2(OH)_2$ structure.

The interpretation of the magnetic properties and magnetic structure require an understanding of the crystal structure, so it is presented in some detail here. The Co–O–Co 2D layers are formed by sharing edges between the $Co1O_6$-*oct* and $Co2O_6$-*oct*. Figure 2 shows the formation of 2D Co–O–Co layer via connectivity between oxygen atoms. In Figure 2, Co1 and Co2 are represented by cyan and blue colors, respectively to highlight the connectivity between $Co1O_6$-*oct* and $Co2O_6$-*oct*. Within the 2D layer, $Co2O_6$-*oct* connects with another $Co2O_6$-*oct* along the *b*-axis by sharing edges via O1 and O2 atoms which creates a Co2–O–Co2 chains along the *b*-axis. The $Co1O_6$-*oct* provides the connectivity between Co2–O–Co2 chains along *c*-axis. For this, each $Co1O_6$-*oct* connects with two $Co2O_6$ sharing edges via O1 and O3. It is interesting to point out that O1 connects two Co2 sites and one Co1 site creating a $Co_3$-triangular unit as shown in Figure 2. In this $Co_3$-triangular unit, O1 serves as the center of the $Co_3$ triangle as a $\mu_3$-oxo atom. At the same time, O1 also connects with H atoms to form the only hydroxide group in the structure. These $Co_3$-triangular units share edges and corners with each other in an alternate fashion creating triangular chains along the *c*-axis. Each triangular chains interconnect with each other via Co2–Co2 corners to form the 2D kagome strip lattice. This specific connection creates alternating triangles and honeycombs running along the *a*-axis which is highlighted in Figure 3a and 3b. Figure 3b highlights the construction of the one-dimensional (1D) kagome strip chain. This 1D kagome strip chain can be considered as a dimensional reduction of the 2D kagome magnetic lattice which can be used to simplify the theoretical aspects of 2D kagome magnetic lattice.

Table 3 summarizes the bond distances and angles of $Na_2Co_3(AsO_4)_2(OH)_2$. The $CoO_6$-*oct* units have a high degree of distortion with Co1–O bond distances ranging from 1.902(1) to 2.1668(1) Å and Co2–O distances range from 2.0407(1) to 2.1310(1) Å. The average Co–O bond distances of $Co1O_6$ and $Co2O_6$ units are 2.1001(2) and 2.0992(1) Å, respectively which are comparable to the expected sum of the Shannon crystal radii, (2.145 Å) for a 6-coordinate high spin $Co^{2+}$ and $O^{2-}$. As shown in Figure 3b, Co–Co distances are 2.798 (Co1–Co1), 2.994 (Co1–Co2) and 3.140 Å (Co2–Co2). The magnetic coupling values for Co are defined as $J_1$, $J_2$ and $J_3$ for these distances, respectively. The nearest distance between the two Co–Co kagome strip layers is 7.3 Å which is *a*/2 (Figure 3c). Since the kagome strip layers do not connect directly via $AsO_4$ groups with one another, we can ignore the interlayer magnetic exchange interactions. Furthermore, the bond angles associated with $J_1$, $J_2$ and $J_3$ exchange interactions are Co1–O1–Co2; 96.63(3)°, Co1–O3–Co2; 88.34(3)°, Co1–O2–Co2; 100.65(3)°, and Co2–O2–Co2; 82.01(2)° which add further complexity to the magnetic properties of $Na_2Co_3(AsO_4)_2(OH)_2$.

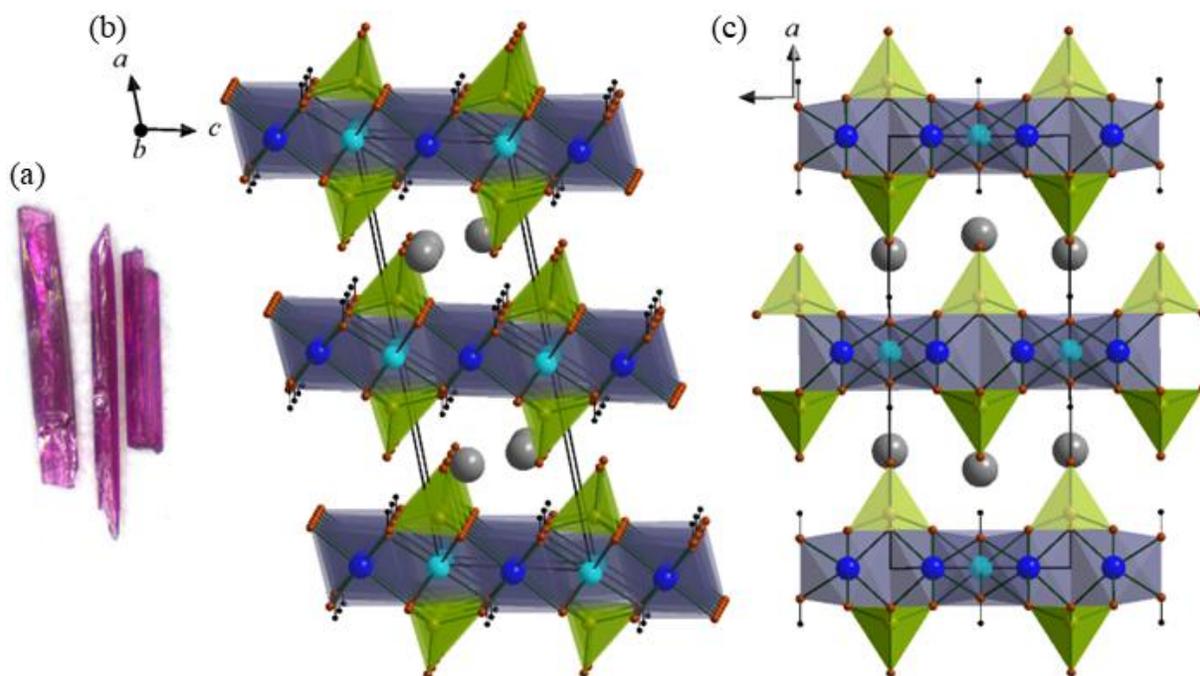

**Figure 1.** (a) Hydrothermally grown Na$_2$Co$_3$(AsO$_4$)$_2$(OH)$_2$ single crystals. (b) Two-dimensional framework of Na$_2$Co$_3$(AsO$_4$)$_2$(OH)$_2$ showing the layered structure along the *b*-axis. (c) Displaying the layered structure along the *c*-axis. The layers are made from CoO$_6$ octahedra and AsO$_4$-terahedra. The Na-ions packed in between the layers. Sodium (Na) atoms are represented in grey, whereas Co1 and Co2 sites are depicted in cyan and blue, respectively. Arsenic (As), oxygen (O), and hydrogen (H) atoms are illustrated in dark yellow, brown, and black, respectively.

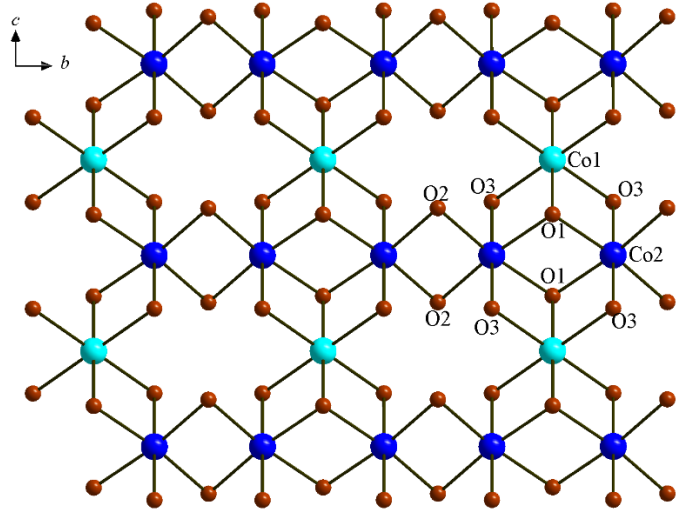

**Figure 2.** Shows the connectivity between $Co1O_6$-*oct* and $Co2O_6$-*oct* within the Co1–O–Co2 2D layer along *bc*-plane.

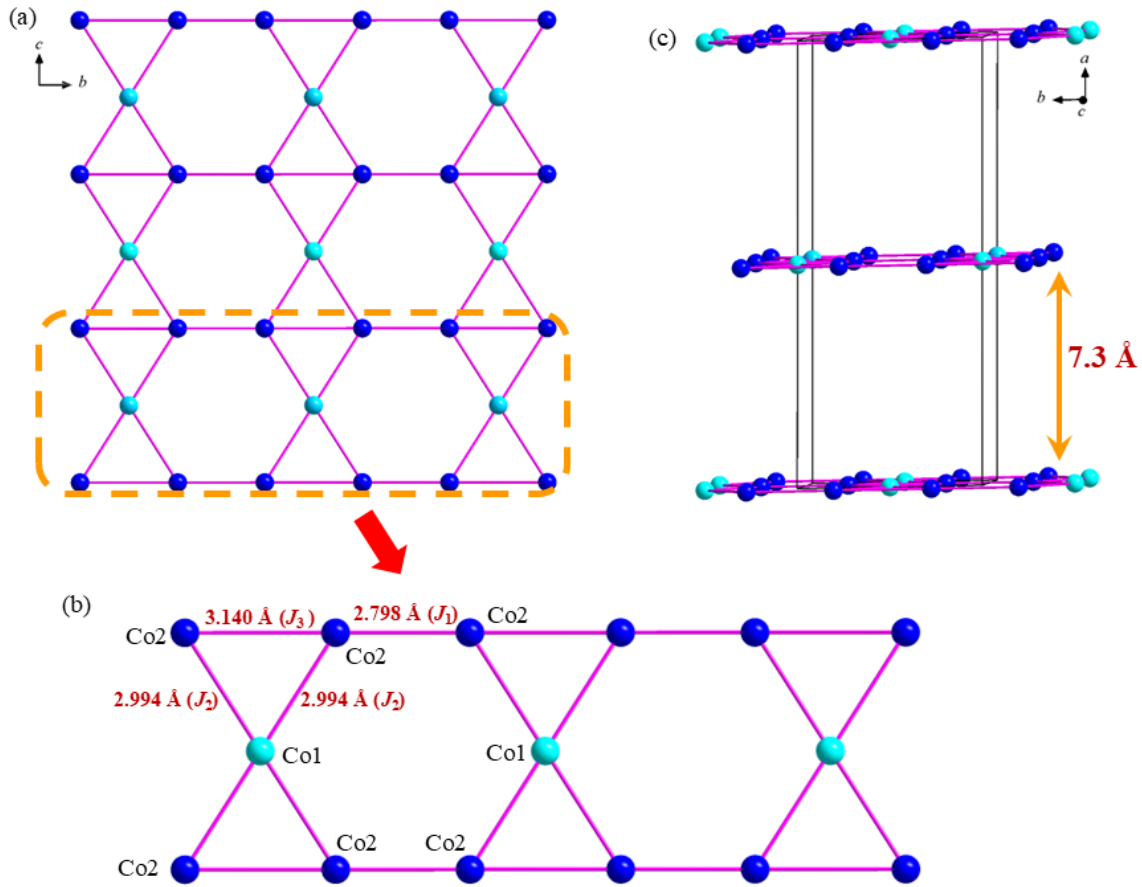

**Figure 3.** Two-dimensional Co–Co kagome strip layer along the *bc*-plane. Here, Co1 and Co2 are shown in two different colors for clarity (Co1-cyan; Co2-blue). (b) One-dimensional kagome strip chain showing different exchange interactions within the chain. (c) Packing of 2D Co–Co kagome strip layers along the *a*-axis. The interlayer distance is around 7.3 Å.

**Magnetic Properties of $Na_2Co_3(AsO_4)_2(OH)_2$**

The magnetic susceptibility data were collected on $Na_2Co_3(AsO_4)_2(OH)_2$ powder sample at 1 kOe applied magnetic field are shown in Figure 4. The magnetic susceptibility has a peak at 14 K indicating long range antiferromagnetic behavior of $Na_2Co_3(AsO_4)_2(OH)_2$ system. In spite of the low dimension of the Co–O–Co magnetic lattice, the susceptibility does not exhibit the broad hump one might expect. This may be attributed to the use of a powder sample for magnetic measurement as opposed to an oriented single crystal. The inverse magnetic susceptibility in the paramagnetic region (150-350 K) was fitted (Figure 4) using the Curie-Weiss model ($M/H = C(T-\theta)$), which produced an effective magnetic moment of 5.3(1) $\mu_B$ per Co and a Weiss temperature of -0.25(2) K. The effective magnetic moment obtained from CW fit is much larger than the spin only value for high-spin $Co^{2+}$ ($3d^7$; $S = 3/2$, 3.87 $\mu_B$), due to the significant orbital contribution from the unquenched orbital moment of $Co^{2+}$ in octahedral environment.[52-55] The low Weiss value is somewhat surprising however, given that Co–O–Co lattice can be considered as a highly frustrated magnetic lattice. This means $Na_2Co_3(AsO_4)_2(OH)_2$ is less frustrated compared to the other reported kagome strip materials. For example, $A_2Cu_5(TeO_3)(SO_4)_3(OH)_4$ ($A$ = Na, K), $\alpha$-$Cu_5O_2(SeO_3)_2Cl_2$ and $Ca_2Mn_3O_8$ show a broad hump in their magnetic susceptibility data at higher temperatures (before long-range magnetic ordering) even in the powder samples suggesting magnetic frustration due to the low-dimensionality of the magnetic lattice.[31-33]

Isothermal magnetization data was determined on the ground single crystals sample up to 90 kOe applied magnetic field and temperatures ranging from 2 to 100 K (Figure 5). At 2 K, the isothermal magnetic data exhibits a plateau, followed by a sharp upturn at approximately 30 kOe.

As the field approaches 90 kOe, magnetization slowly increases in tendency towards saturation or another field-induced transition. The magnetization value of the plateau state (approximately 0.25 $\mu_B$/Co) corresponds to ~1/9 of the total saturation value of $Co^{2+}$ ($S = 3/2$, 3.8 $\mu_B$/Co). This field induced transition at 30 kOe could be a spin-flop like transition where Co1 and Co2 moments arrange perpendicular to the applied magnetic field direction. With the continuous increase of the magnetic field these spins can gradually rotate towards the applied magnetic field becoming a ferromagnetic phase.[56-59]

The temperature evolution of magnetic susceptibility was measured up to 90 kOe from 2 to 40 K, across the long-range ordering transition (Figure 5b). The temperature-dependent magnetic susceptibilities of $Na_2Co_3(AsO_4)_2(OH)_2$ in different applied magnetic fields (Figure 5b) also support the spin flop transition that was observed in isothermal magnetization data (Figure 5a). The low temperature downturn becomes broader, and the $T_N$ gradually shifts to lower temperatures with the increase of applied magnetic fields (Figure 5b). At 50 kOe, the magnetic susceptibility does not exhibit a round maximum. The development of a broader maximum and shift of $T_N$ to lower temperatures could be due to the spin flop transition as seen in the isothermal magnetization data. Overall $Na_2Co_3(AsO_4)_2(OH)_2$ can be identified as a canted antiferromagnetic structure with a spin flop type transition. This is possible since $Na_2Co_3(AsO_4)_2(OH)_2$ possess two crystallographically different magnetic sites (Co1 and Co2) in the Co–O–Co magnetic lattice.[57-60]

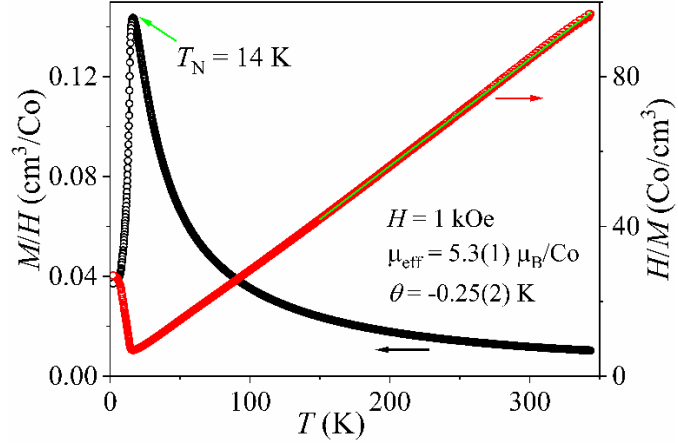

**Figure 4.** Magnetic susceptibility ($\chi$) and inverse magnetic susceptibility ($1/\chi$) as a function of temperature measured at 1 kOe applied magnetic field. Curie-Weiss fit (150-350 K) is shown in green line.

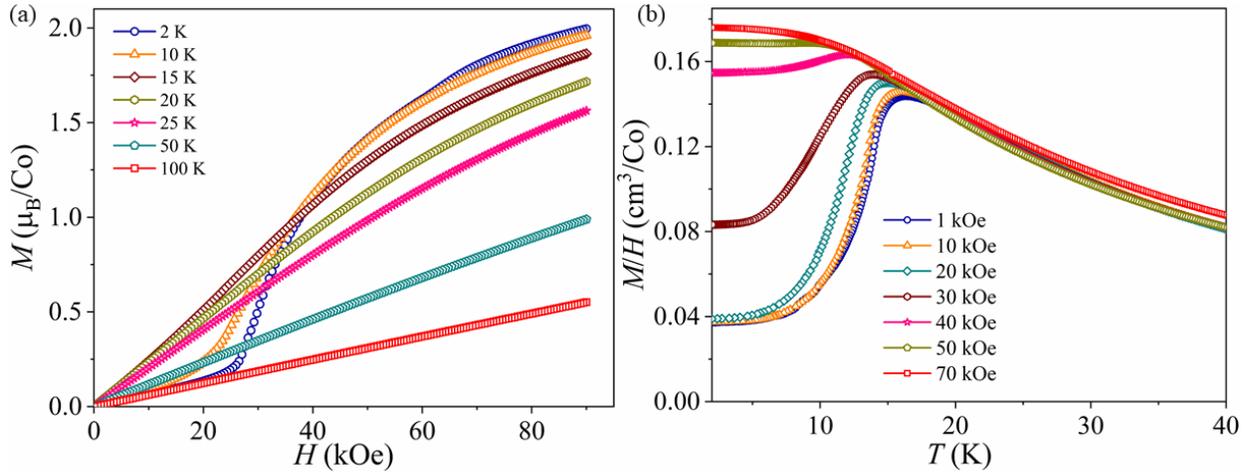

**Figure 5.** (a) Isothermal magnetization data collected in different temperatures (2-100K). (b) Magnetization measured at lower temperatures (2-40 K) in different applied magnetic fields (1-70 kOe).

**Magnetic Structure of Na$_2$Co$_3$(AsO$_4$)$_2$(OH)$_2$**

Motivated by its intriguing magnetic properties, we used neutron powder diffraction to determine the magnetic structure of Na$_2$Co$_3$(AsO$_4$)$_2$(OH)$_2$ and compare it with other kagome strip structures. An initial NPD pattern was collected at 40 K using the wavelength of 1.54 Å which covers a broader $Q$-range, (Figure 6a). At 1.5 K and 40 K, NPD was performed using the 2.41 Å wavelength to access lower $Q$-range for an improved magnetic structure determination (Figure 6b, c). Crystal and magnetic structure refinements were carried out using FullProf Suite[47] and the refined crystal structure parameters are given in Table SI1. In comparison to the room temperature X-ray diffraction data (Table 1), the crystallographic parameters obtained by NPD refinements at 40 K indicate little change in the lattice parameters. Due to the smaller thermal expansion at lower temperatures, it is expected that the unit cell parameters will be smaller at 1.5 and 40 K. The powder diffraction data at 1.5 K clearly indicates the presence of magnetic Bragg peaks, compared to the 40 K data, due to the long-range order at $T_N$ = 14 K. Figure 6b shows the comparison between the 40 and 1.5 K NPD patterns. No additional peaks or peak splitting appeared at higher $Q$ below $T_N$ suggesting no structural phase transition. The magnetic peaks were indexed using the "$k$-search" program included in the FullProf Suite. The additional reflections can be indexed using the propagation vector $\boldsymbol{k}$ = (0.5,0.5,0.5), which doubles the original unit cell along all three crystallographic directions. Based on the $\boldsymbol{k}$ = (0.5, 0.5, 0.5) and non-magnetic space group $C2/m$, representational analysis was performed using the *SARAh*-program.[48,49] Additionally, the magnetic space group was determined using the MAXMAGN program at the Bilbao Crystallographic Server.[50] According to the representational analysis, the decomposition of the magnetic representations for Co1 can be written as $\Gamma_{mag} = 0\Gamma_1^1 + 3\Gamma_2^1$ and for Co2 can be written as $\Gamma_{mag} = 3\Gamma_1^1 + 3\Gamma_2^1$. The basis vectors for the space group $C2/m$ with $\boldsymbol{k}$ = (0.5,0.5,0.5) are summarized in Table SI2. The best fit was achieved using $\Gamma_2$ and $\Gamma_1$ for Co1 and Co2, respectively. The $\Gamma_2$ irreducible representation allows Co1 to have moment along all three crystallographic directions while arranging antiferromagnetically along the $bc$-plane. The $\Gamma_1$ irreducible representation allows Co2 to have moment along all three crystallographic directions while the next nearest neighbors aligned antiferromagnetically along the $b$-axis. This model allows us to obtain a better fit for the NPD at 1.5 K. The magnetic structure in this configuration corresponds to the magnetic space group of $P_s$-1 (#2.7), with a magnetic unit cell of $2a \times 2b \times 2c$. Full details of the magnetic space group are given in Table SI3. This magnetic space group was the only maximal group that allowed both Co ions to have non-zero moments. The moments for each Co ion were not constrained by symmetry and were allowed to freely vary. The refined magnetic moments are $m_{Co1}$ = 4.3(1)$\mu_B$ ($m_a$ = 2.51(2)$\mu_B$; $m_b$ = -0.27(1)$\mu_B$; $m_c$ = -3.01(1)$\mu_B$) and $m_{Co2}$ = 2.87(7)$\mu_B$ ($m_a$ = 0.51(1)$\mu_B$; $m_b$ = 2.83(7)$\mu_B$ and $m_c$ = 0.16(1)$\mu_B$). Even though both Co1 and Co2 have moments along all three crystallographic axes, the Co1 magnetic moments mostly lie on the *ac*-plane while the Co2 moments are mainly aligned along the *b*-axis. Figure 7 shows the magnetic model from the best fit of the NPD pattern, drawn using the expanded magnetic unit cell (2*a*, 2*b*, 2*c*). As shown in Figure 7a, the Co1 moments are mostly directed along the *[101]* direction, with, the nearest neighbor spins coupled antiferromagnetically along the *b*-axis. The Co2 moments lie mostly along the *b*-axis with a very small out of plane (*a*-axis) canting. As noted, the Co2 moment along the *a*- and *c*-axes are very small and closer to zero within the reported errors. Nearest neighbor Co2 sites in the triangle ($J_3$) couple antiferromagnetically while nearest neighbor Co2 sites within the honeycomb ($J_1$) coupled ferromagnetically. The field induced transition could be due to the flipping the direction of the Co1 moment towards Co2 or vice versa under the applied magnetic field. The anisotropic magnetic properties will be essential to confirm this. Figure 7b displays the packing of 2D kagome strip magnetic lattices along the *a*-axis. The moment of Co1 site is 4.3(1)$\mu_B$ which is higher than the calculated moment of $Co^{2+}$ ($S$ = 3/2; 3.4$\mu_B$). In contrast, the moment of Co2 is relatively smaller 2.87(7) $\mu_B$ than the expected moment of $Co^{2+}$. The difference between magnetic moments of Co1 and Co2 is expected due to the triangular arrangement of the magnetic lattice. A similar behavior was observed in our previous studies with two crystallographic ions that are magnetic, CsCo$_2$(MoO$_4$)$_2$(OH) and K$_2$Co$_3$(MoO$_4$)$_3$(OH)$_2$. The relatively higher magnetic moment of Co1 is indeed unexpected. We emphasize that the neutron refinements were thoroughly examined to rule out artifacts or over-fitting, including checks on the magnetic structure model, scale factor, and background. In addition, the full refinement parameters are provided in the Supporting Information. Nevertheless, we cannot rule out that limitations in the magnetic structure refinement or experimental factors inherent to neutron powder diffraction may contribute to the discrepancy observed between the Co1 and Co2 moment values.

The magnetic structure of Na$_2$Co$_3$(AsO$_4$)$_3$(OH)$_2$ is in contrast with the previously reported Mn$^{4+}$-based kagome strip material, Ca$_2$Mn$_3$O$_8$ where the moment exclusively lies along the *a*-axis (out of plane).[33] The magnetic structure of the kagome strip compound, α-Cu$_5$O$_2$(SeO$_3$)$_2$Cl$_2$ has been reported as well, and possesses a complicated magnetic structure due to the presence of three different Cu sites in two different geometries.[32]

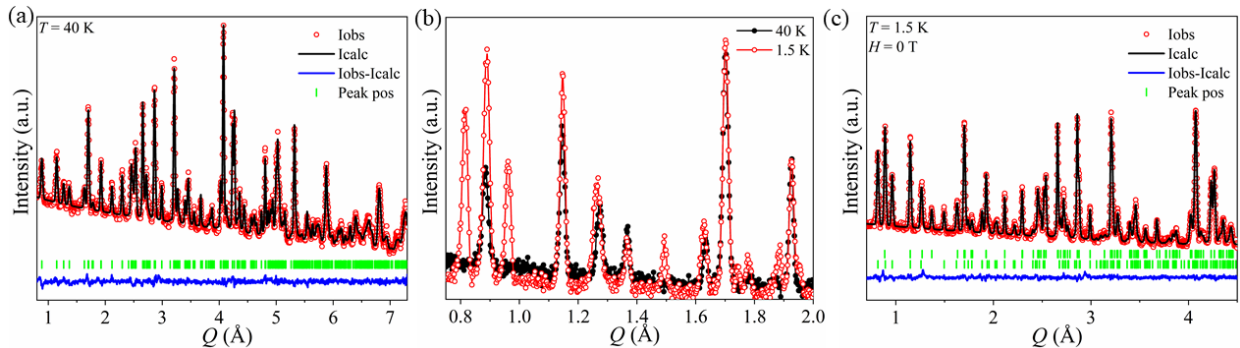

**Figure 6.** (a) Rietveld plot of the neutron powder diffraction data collected at 40 K using 1.54 Å. (b) Overlay of NPD patterns of 40 K and 1.5 K collected using 2.41 Å wavelength. (c) Rietveld plot of the neutron powder diffraction data collected at 1.5 K using 2.41 Å wavelength.

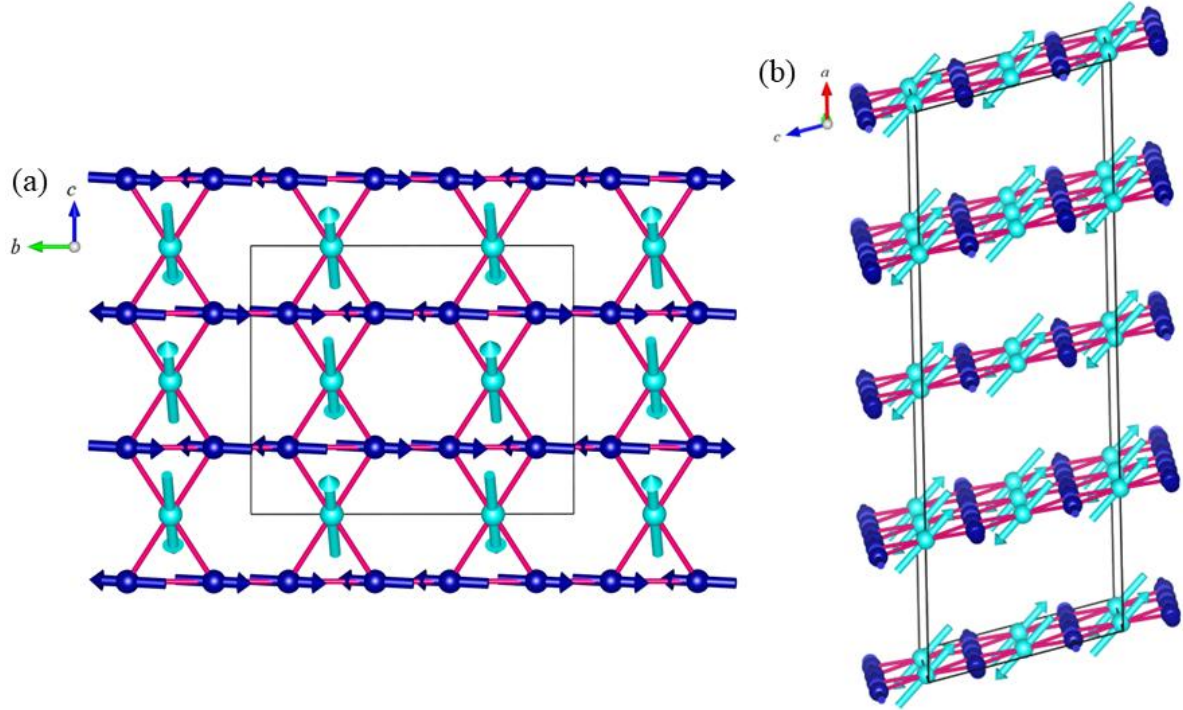

**Figure 7.** A view of the magnetic structure. (a) The 2D kagome strip magnetic lattice along the *bc*-plane. (b). The 2D kagome strip magnetic lattices are coupled antiferromagnetically along the *a*-axis within the unit cell. The magnetic structure is drawn using its magnetic unit cell which is different from the nuclear cell due to the ***k***-vector (0.5, 0.5, 0.5).

**Inelastic Neutron Scattering and Exchange Parameters of $Na_2Co_3(AsO_4)_2(OH)_2$**

Inelastic neutron scattering (INS) measurements were performed on a powder sample of $Na_2Co_3(AsO_4)_2(OH)_2$ using HYSPEC at 5 K to investigate magnetic excitations associated with the long-range magnetic order of $Co^{2+}$ moments in its two-dimensional kagome-strip lattice. Data were collected using incident energies of $E_i = 50$ meV and 20 meV. The $E_i = 50$ meV spectra reveals two sharp excitations at 26 meV and 32 meV (Figure 8a), while the $E_i = 20$ meV data shows two lower-energy magnetic excitation bands centered around 5 meV and 10 meV (Figure 8b). $Na_2Co_3(AsO_4)_2(OH)_2$ contains two crystallographically distinct $Co^{2+}$ sites, which can naturally lead to multiple magnetic excitation modes. In an octahedral crystal field, $Co^{2+}$ ($3d^7$) ions have an effective spin $S = 3/2$ and orbital angular momentum $L_{eff} = 1$. Spin-orbit coupling (with strength $\lambda$) splits the ground state into a $J_{eff} = 1/2$ ground state and excited $J_{eff} = 3/2$ and $5/2$ states, separated by energy gaps of $3/2\,\lambda$ and $5/2\,\lambda$, respectively. The observed higher-energy bands at 26 meV and 32 meV are consistent with spin-orbit excitations from the $J_{eff} = 1/2$ ground state to $J_{eff} = 3/2$, arising from the two inequivalent $Co^{2+}$ sites. Their magnetic origin is confirmed by the decreasing intensity with increasing momentum transfer. Similar crystal electric field (CEF) excitations have been reported in other $Co^{2+}$ compounds such as $NaCaCo_2F_7$ and $CoTiO_3$.[60,61]

At 2 K, two nearly flat magnetic excitation bands were observed at 5 meV and 10 meV. To model these low-energy modes, we performed linear spin wave calculations using the **Sunny.jl** package, based on a minimal Heisenberg Hamiltonian incorporating three nearest-neighbor exchange interactions: $J_1$, $J_2$, and $J_3$. Here, $J_1$ and $J_3$ represent couplings within the bond-alternating Co2–Co2 chains, while $J_2$ describes the interaction between Co1 and Co2 sites, as shown in Figure 3b. To reflect the results of magnetic structure refinement, we included single-ion anisotropy terms $D_1$ and $D_2$, aligning Co1 spins along the $c$-axis and Co2 spins along the $b$-axis. These strong anisotropies enforce orthogonality between Co1 and Co2 spins. As a result, the $J_2$ exchange term, which involves perpendicular spins, contributes zero energy since $S_1 \cdot S_2 = 0$. For computational simplicity—without compromising physical accuracy—we set $J_2 = 0$. This simplification improved numerical stability and preserved the directional constraints of the system, enabling powder-averaged simulations that reproduce the experimental INS spectra with high fidelity. Taking into account of several parameters ($J_1$, $J_3$, $D_1$ and $D_2$) and significant anisotropic information loss in powder average INS data, we use an iterative optimization procedure to narrow down the best model which is consistent with the observed INS data and the magnetic structure of $Na_2Co_3(AsO_4)_2(OH)_2$. To assess the uniqueness of our solutions and ensure no viable parameter regions were overlooked, we conducted a systematic search of the full parameter space for ($J_1$, $J_3$, $D_1$, $D_2$) within the range [-12, 12] meV, with $J_2$ fixed at 0. After testing various configurations, we identified several parameter sets that provided relatively good fits to INS data, particularly in reproducing the two flat modes at 5 and 10 meV. These optimal configurations cluster into five distinct phases, labeled #1 through #5, as shown in Figure 8a. Moreover, this comprehensive search revealed that $D_1$ is restricted to only two discrete values: -5 meV for phases #1 and #2, and -2.5 meV for phases #3 – #5. The agreement with the experimentally refined spin configurations $m_{exp}$ was confirmed for all phases with $mean(|m_{simul} - m_{exp}|^2) < 0.1$. The goodness of fit to the INS data was evaluated using the relative squared deviation (RSD), $\frac{\Sigma |S_{simul}(Q,E) - S_{exp}(Q,E)|^2}{\Sigma |S_{exp}(Q,E)|^2}$. Figure 9 shows the possible solutions sets in the ($J_1$, $J_3$) and ($D_2$, $J_3$) parameter space, with points color-coded by phase number (grey, red, green, blue, and orange for phases #1 – #5, respectively). Here, point size and transparency reflect the RSD values, where larger, less transparent points indicate better fits. The best parameter set within each phase is marked with an "**x**", with phase #1 achieving the global minimum of RSD = 0.358. The corresponding simulated $S(Q, E)$ spectrum for the globally optimal phase #1 is shown in Figure 9b, while Figure 9c presents the $S(Q, E)$ spectra for the best parameter sets from #2 to #5 which exhibit slightly higher RSD values.

The complete coordinates ($J_1$, $J_2$, $J_3$, $D_1$, $D_2$) and RSD values for the optimal parameter sets in each phase are listed in Table 4. While phase #1 achieves the best RSD value, both the quantitative RSD differences and the qualitative comparison of the simulated $S(Q, E)$ spectra (Figure 9b-c) reveal relatively small variations between phases, making it difficult to definitively exclude alternative solutions based solely on the current INS data. Further measurements—such as on single crystals or complementary experimental techniques—would be necessary to further constrain the parameter space and distinguish between these competing solutions to achieve a more definitive determination of the magnetic exchange parameters. Additionally, the lack of spin dynamic studies on Kagome strip materials prevents us from comparing our results with those of similar systems. The inelastic neutron scattering data of $Ca_3Mn_2O_8$ has been reported, however no detailed model is provided. The INS data of $Ca_2Mn_3O_8$ data shows some diffuse modes below and above the $T_N$, suggesting some spin fluctuations because of the low dimensionality of the magnetic lattice.

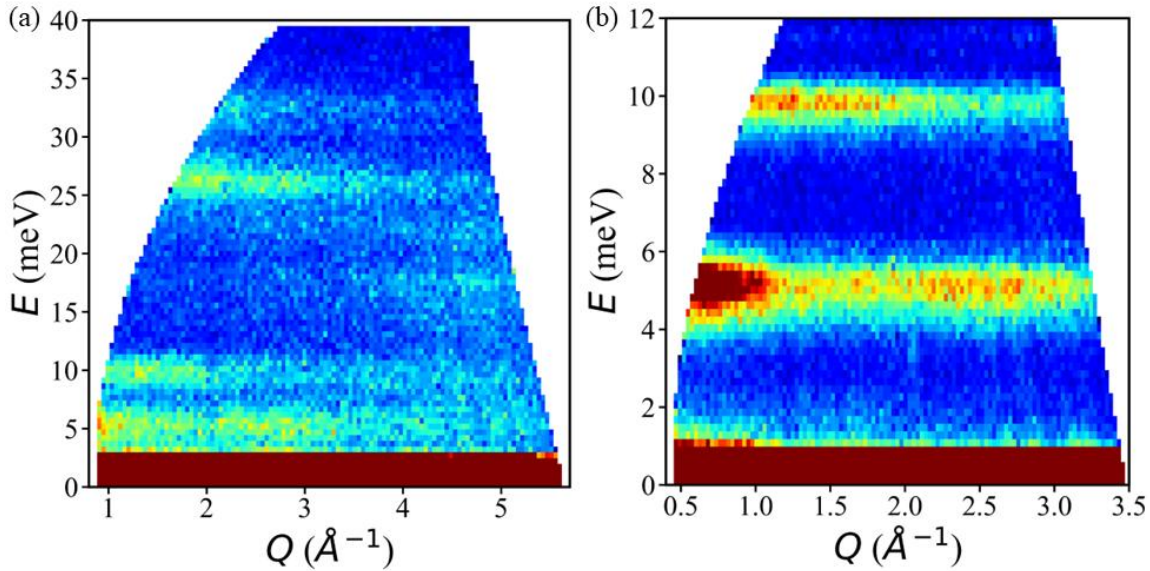

**Figure 8.** (a) INS data with $E_i$ = 50 meV, showing CEF excitations at 26 meV and 32 meV. (b) INS data with $E_i$ = 20 meV, showing magnetic excitations at 5 meV and 10 meV.

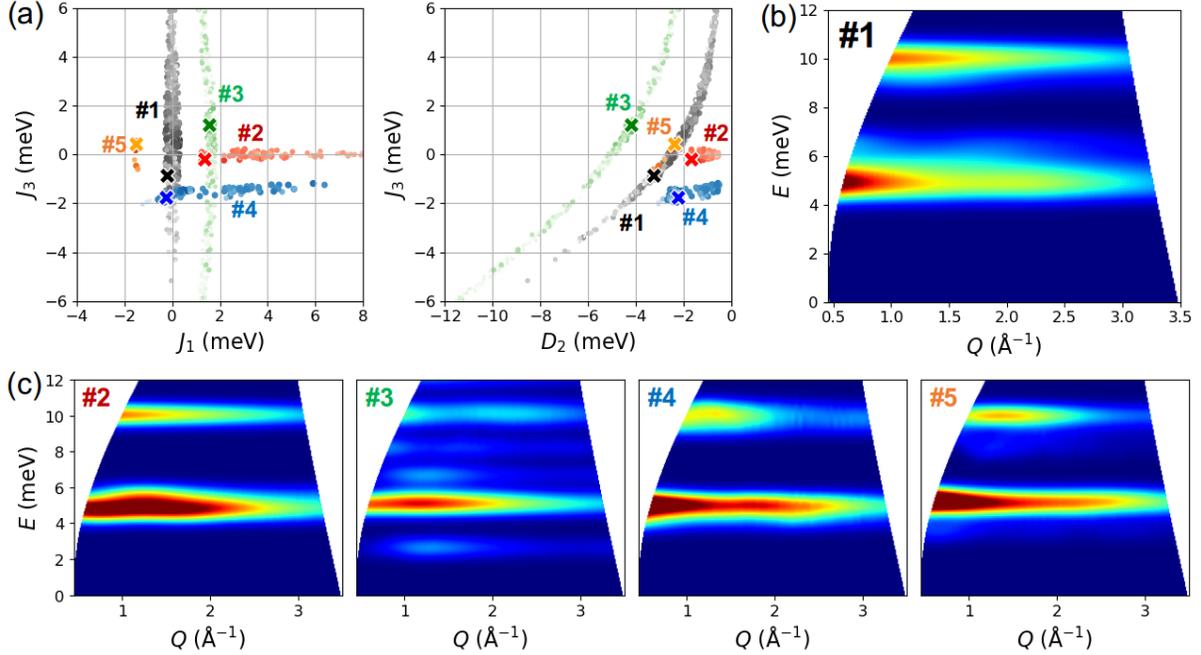

**Figure 9.** (a) Scatter plot of parameter sets in the ($J_1$, $J_3$) (left) and ($D_2$, $J_3$) (right) parameter space. Points are color-coded by phase number (#1 – #5): grey, red, green, blue, and orange. Point size and transparency reflect RSD, where larger size and lower transparency indicate better RSD values. For each phase, the best point (lowest RSD) is marked with "**x**". For all phases, $J_2$ is fixed to 0, and $D_1$ takes discrete values: –5 meV for phases #1 and #2, and –2.5 meV for phases #3 – #5. (b) Simulated $S(Q, E)$ spectrum for the best point in phase #1, which represents the global optimum (lowest RSD). (c) Simulated $S(Q, E)$ spectra for the best points in phases #2 – #5, which have slightly higher RSD values.

**Table 4** Best-fit parameter sets ($J_1$, $J_2$, $J_3$, $D_1$, $D_2$) and corresponding RSD values for phases #1– #5. Note: The $J_2$ term contributes no energy because the Co1–Co2 spins are orthogonal ($S_1 \cdot S_2 \approx 0$); therefore, it is fixed to zero (see the text for full details).

Best parameter sets for each phase: ($J_1$, $J_2$, $J_3$, $D_1$, $D_2$) coordinates and corresponding RSD values for phases #1 – #5.

| Phase | RSD | $J_1$ (meV) | $J_2$ (meV) | $J_3$ (meV) | $D_1$ (meV) | $D_2$ (meV) |
|---|---|---|---|---|---|---|
| #1 | 0.358 | -0.23 | 0 | -0.86 | -5.0 | -3.26 |
| #2 | 0.468 | 1.34 | 0 | -0.18 | -5.0 | -1.67 |
| #3 | 0.456 | 1.53 | 0 | 1.23 | -2.5 | -4.20 |
| #4 | 0.379 | -0.24 | 0 | -1.74 | -2.5 | -2.22 |
| #5 | 0.375 | -1.53 | 0 | 0.46 | -2.5 | -2.37 |

**Conclusions**

We report the synthesis and comprehensive magnetic analysis of $Co^{2+}$ ($S_{eff}$ = ½) kagome strip compound, $Na_2Co_3(AsO_4)_2(OH)_2$. In recent years, theoretical studies on kogome strip magnetic materials have predicted the presence of multiple magnetization plateaus that are quantum in nature. The focus of our research has been on the study of static order and spin dynamics of this novel class of materials to better understand the inherited quantum properties. $Na_2Co_3(AsO_4)_2(OH)_2$ crystalizes in a monoclinic crystal system with a nearly perfect kagome strip magnetic lattice with a larger interlayer separation ~7 Å. Bulk magnetic properties reveal an antiferromagnetic transition at $T_N$ = 14 K that shows some field dependence. At 1.5 K, neutron powder diffraction was used to characterize the zero-field magnetic structure which can be described using the *k*-vector (0.5, 0.5, 0.5). In the proposed magnetic structure, the Co1 moments are mostly directed along the [101] direction, with, the nearest neighbor spins coupled antiferromagnetically along the *b*-axis. The Co2 moments lie mostly along the *b*-axis with a very small out of plane (*a*-axis) canting. Inelastic neutron scattering confirms the presence of crystal field excitations which supports the presence of $J_{eff}$ =1/2 state due to the unquenched spin-orbital coupling. The spin wave spectrum collected at $E_i$ = 20 meV comprises two flat bands centered at 5 meV and 10 meV. A simplified Heisenberg Hamiltonian model that includes three nearest neighbors exchange interactions and single ion anisotropy is sufficient to describe the magnetic excitations. We conclude that our work illustrates the complexity of magnetism arising from kagome strip magnetic lattices. To gain a deeper understanding of this system, it will be necessary to conduct further magnetic and neutron scattering studies using single crystals. The aim of our study is to inspire the search for other $Co^{2+}$ and $Cu^{2+}$ based kagome strip magnetic structures.

**Author Contributions**

D. S. Liurukara lead the writing and data analysis. E. D. Williams performed sample synthesis and structure characterization. S. Calder performed neutron powder diffraction experiment and data analysis. V. O. Garlea performed INS experiment and data analysis. T. Chen performed INS data analysis. C. C. Buchanan and D. A. Gilbert perform bulk magnetic property measurements. J. W. Kolis and D. A. Tennant involved in writing and all co-authors made comments on the paper.

**Conflicts of Interest**

There are no conflicts to declare.

**Acknowledgements**

The research at the Oak Ridge National Laboratory (ORNL) is supported by the U.S. Department of Energy (DOE), Office of Science, Basic Energy Sciences (BES), Materials Sciences and Engineering Division. This research used resources at the High Flux Isotope Reactor and Spallation Neutron Source, DOE Office of Science User Facilities operated by ORNL. The beam time was allocated to HB-2A (POWDER) and HYSPEC (BL14B) on proposal numbers IPTS-32168 and 33656.] The synthesis, crystal growth and X-ray diffraction were supported by NSF award DMR – 2219129. Magnetic measurements taken at the University of Tennessee were supported by the DOE Early Career Program, Award DE-SC0021344.

This manuscript has been coauthored by employees of ORNL, which is managed by UT-Battelle, LLC, under Contract No. DE-AC05-00OR22725 with the U.S. Department of Energy (DOE). The U.S. government retains and the publisher, by accepting the article for publication, acknowledges that the U.S. government retains a nonexclusive, paid-up, irrevo-cable, worldwide license to publish or reproduce the published form of this manuscript, or allow others to do so, for U.S. government purposes. DOE will provide public access to these results of federally sponsored research in accordance with the DOE Public Access Plan.[62]

**Data Availability Statement**

The data supporting this article has been included as part of the manuscript supporting information. CCDC 2484978 contains the supporting crystallographic data for this paper. These data can be obtained free of charge via www.ccdc.cam.ac.uk/data_request/cif (accessed on 25 August 2022), or by emailing data_request@ccdc.cam.ac.uk, or by contacting The Cambridge Crystallographic Data Centre, 12 Union Road, Cambridge CB2 1EZ, UK; fax: +44-1223-336033.